\documentclass[twocolumn]{aastex62}
\usepackage{epsfig}
\usepackage{graphicx}
\usepackage{float}
\usepackage{amsmath}
\usepackage{color}
\usepackage{amssymb}
\usepackage{amsfonts}
\usepackage{units}
\usepackage{amsmath,bm}
\usepackage{amssymb}
\usepackage{amsfonts}
\usepackage{subfigure}
\usepackage{etoolbox} 

\newcommand{\frb}{FRB 180916B}

\begin{document}

\title{Periodic Activities of Repeating Fast Radio Bursts from Be/X-ray Binary Systems}

\author{Qiao-Chu Li}
\affil{School of Astronomy and Space Science, Nanjing University, Nanjing 210023, China}
\affil{Key laboratory of Modern Astronomy and Astrophysics (Nanjing University), Ministry of Education, Nanjing 210023, China}

\author{Yuan-Pei Yang}
\affil{South-Western Institute for Astronomy Research, Yunnan University, Kunming 650500, China; ypyang@ynu.edu.cn}

\author{ F. Y. Wang}
\affil{School of Astronomy and Space Science, Nanjing University, Nanjing 210023, China}
\affil{Key laboratory of Modern Astronomy and Astrophysics (Nanjing University), Ministry of Education, Nanjing 210023, China}

\author{Kun Xu}
\affil{School of Astronomy and Space Sciences, University of Chinese Academy of Sciences, Beijing, China}
\affil{Key Laboratory of Optical Astronomy, National Astronomical Observatories, Chinese Academy of Sciences, Beijing, China}

\author{Yong Shao}
\affil{School of Astronomy and Space Science, Nanjing University, Nanjing 210023, China}
\affil{Key laboratory of Modern Astronomy and Astrophysics (Nanjing University), Ministry of Education, Nanjing 210023, China}

\author{Ze-Nan Liu}
\affil{School of Astronomy and Space Science, Nanjing University, Nanjing 210023, China}
\affil{Key laboratory of Modern Astronomy and Astrophysics (Nanjing University), Ministry of Education, Nanjing 210023, China}

\author{Zi-Gao Dai}
\affil{Department of Astronomy, School of Physical Sciences, University of Science and Technology of China, Hefei 230026, China; daizg@ustc.edu.cn}
\affil{School of Astronomy and Space Science, Nanjing University, Nanjing 210023, China}

\begin{abstract}
The frequency-dependent periodic active window of the fast radio burst FRB 180916.J0158+65 (FRB 180916B) was observed recently.
In this Letter, we propose that a Be/X-ray binary (BeXRB) system, which is composed of a neutron star (NS) and a Be star with a circumstellar disk, might be the source of a repeating FRB with periodic activities, and apply this model to explain the activity window of FRB 180916B.
The interaction between the NS magnetosphere and the accreted material  results in evolution of the spin period and the centrifugal force of the NS, leading to the change of the stress in the NS crust.
When the stress of the crust reaches the critical value, a starquake occurs and further produces FRBs. The interval between starquakes is estimated to be a few days that is smaller than the active window of FRB 180916B. When the NS moves out of the disk of the Be star, the interval between starquakes becomes much longer than the orbital period, which corresponds to the non-active phase.
In this model, due to the absorption of the disk of the Be star, a frequency-dependent active window would appear for the FRBs, which is consistent with the observed properties of FRB 180916B. 
And the contribution of dispersion measure (DM) from the disk of the Be star is small.
In addition, the location of FRB 180916B in the host galaxy is consistent with a BeXRB system.

\end{abstract}

\section{Introduction}

Fast radio bursts (FRBs) are a kind of millisecond-duration radio transients with extremely high brightness temperatures \citep[for reviews, see][]{Petroff19, Cordes19,Platts19,zha20b,Xiao21}. So far, hundreds of FRBs have been detected \citep[e.g.,][]{lor07, tho13, spi16, cha17,ban19, rav19,pro19, mar20}
and a part of them show repeating behaviors \citep[e.g.,][]{spi14, chi19, kum19, luo20}.
Recently, a Galactic FRB, FRB 200428, \citep{boc20,chi20b} was discovered from the magnetar SGR J1935+2154, which is associated with an X-ray burst \citep{mer20,li21,tav21,rid21}. This indicates that magnetars are at least one origin of FRBs, while the radiation mechanism of FRBs might be still unclear \citep[e.g.,][]{lyu14, geng15, dai16, wax17, kum17, Lyutikov2017, zha17, kat18, yan18, yan21, met19, Wadiasingh2019, dai20, gen20, kum20, lu20, yan20, bel20, wu20, zha20b, xia20, yu21}.

Remarkably, there are two repeating sources showing periodic activities: FRB 180916.J0158+65 (FRB 180916B), with a 16.35-day periodic activity and a 5-day activity window \citep{Chime/Frb2020}, and FRB 121102, with a possible longer period $\sim160\,\unit{day}$ \citep{raj20, cru21} and an active window about $\sim100~{\rm day}$. Also, possible periodic activities of soft gamma-ray repeaters of SGR 1806-20 and SGR 1935+2154 were found to be about $398.2$ days \citep{zha21} and $237$ days \citep{zou21}, respectively. Several scenarios have been proposed to explain the periodic activities of FRBs  \citep[see][]{zha20}: first, the FRB source is in a binary system with the observed period corresponding to the orbital period of this binary system \citep{dai16, sma19, iok20, lyu20, dai20b, gu20, mot20, dec21, kue21, den21, wad21}; second, the FRB is generated by the NS magnetosphere, and the observed periodic activities are due to the precession of the emitting region \citep{lev20, yangh20, zan20, ton20, lidz21, sri21}; third, the observed period is due to the rotation of a very slow NS \citep{ben20,xu21}.

Recently, FRB 180916B was found to be $\sim 250\,\unit{pc}$ offset from the brightest region of the nearest young stellar clump in its host galaxy, which is suggested to be the birth place of FRB 180916B \citep{ten21}.
It would take $800\,\unit{kyr}$ to $7\,\unit{Myr}$ for FRB 180916B to cross the observed distance from this presumed birth place, which is in tension with scenarios that invoke young ($\lesssim 10\,\unit{kyr}$) magnetars formed in core collapse supernovae \citep{ten21}.
However, such a large offset might be well consistent with the case of the high mass X-ray binaries \citep[HMXBs,][] {bod12}.
In particular, \cite{Pleunis21} recently found that FRB 180916B is preferred to be from an interacting binary featuring a neutron star (NS) and high-mass companion, which accounts for most observational properties of this FRB. Therefore, the relation between FRBs and HMXBs should be extensively investigated.

Notably, there is a subtype of HMXB called Be/X-ray binary (BeXRB), whose typical orbital period \citep[$\sim 10$\,day to 1\,yr;][]{oka02} is similar to the period of \frb. 
Meanwhile, the NS with strong magnetic field \citep[$\sim 10^{12}-10^{13}$\,G;][]{tau06} in the 
BeXRB system could also be the central engine of FRBs.
Recently, \cite{zhanggao20} used the population synthesis method to study in detail the properties of the companion star, aiming to explain the periodicity of FRB 180916B. They found that the companion star is most likely to be a B-type star.
The classical Be star is a main sequence B star, which rotates so fast that an outwardly diffusing gaseous, dust-free Keplerian disk is formed \citep{oka01}.
BeXRBs can exhibit X-ray radiation outbursts that last from a week to several months. One type of outbursts is called normal outbursts or Type \uppercase\expandafter{\romannumeral1} outbursts, which are characterized by X-ray luminosity of $L_{\text{X}} \sim 10^{36}-10^{37} \,\unit{erg\, s^{-1}}$, periodically peaking at or close to the periastron, and covering $0.2-0.3\,P_{\text{orb}}$ of the orbital period \citep[e.g.,][]{rei11}.
The origin of Type \uppercase\expandafter{\romannumeral1} outbursts is usually explained by the accretion of the NS from the Be star disk, when the NS is approaching the Be star and closes enough near the periastron \citep[e.g.,][]{oka01}.

Up to now, with detections of FRB 180916B from $110\,\unit{MHz}$ to $1765\,\unit{MHz}$ \citep{pas20, san20, Chime/Frb2020, mart20, agg20}, the period is confirmed to be $16.29_{-0.17}^{+0.15}\,\unit{day}$, the active window is $6.1\,\unit{day}$, which peaks earlier at higher frequencies \citep{pas20}.
In this Letter, we propose that a BeXRB system could be the source of a repeating FRB with periodic activity.
We discuss the triggering mechanism of FRBs in a BeXRB, and predict the waiting time and energy in Section \ref{sec2.1}.
We use the BeXRB model to explain frequency-dependence active window of FRB 180916B in Section \ref{sec2.2}.
We calculate the dispersion measure (DM) contribution from the disk of the Be star and make a comparison with the observations of FRB 180916B in Section \ref{sec2.3}.
Finally,  we make some discussion and conclusions in Section \ref{sec3}.

\section{Repeating FRBs from a BeXRB system}\label{sec2}

\begin{figure}[]
\centering
\subfigure[]{
\label{Fig.sub.1}
\includegraphics[width=0.46\textwidth]{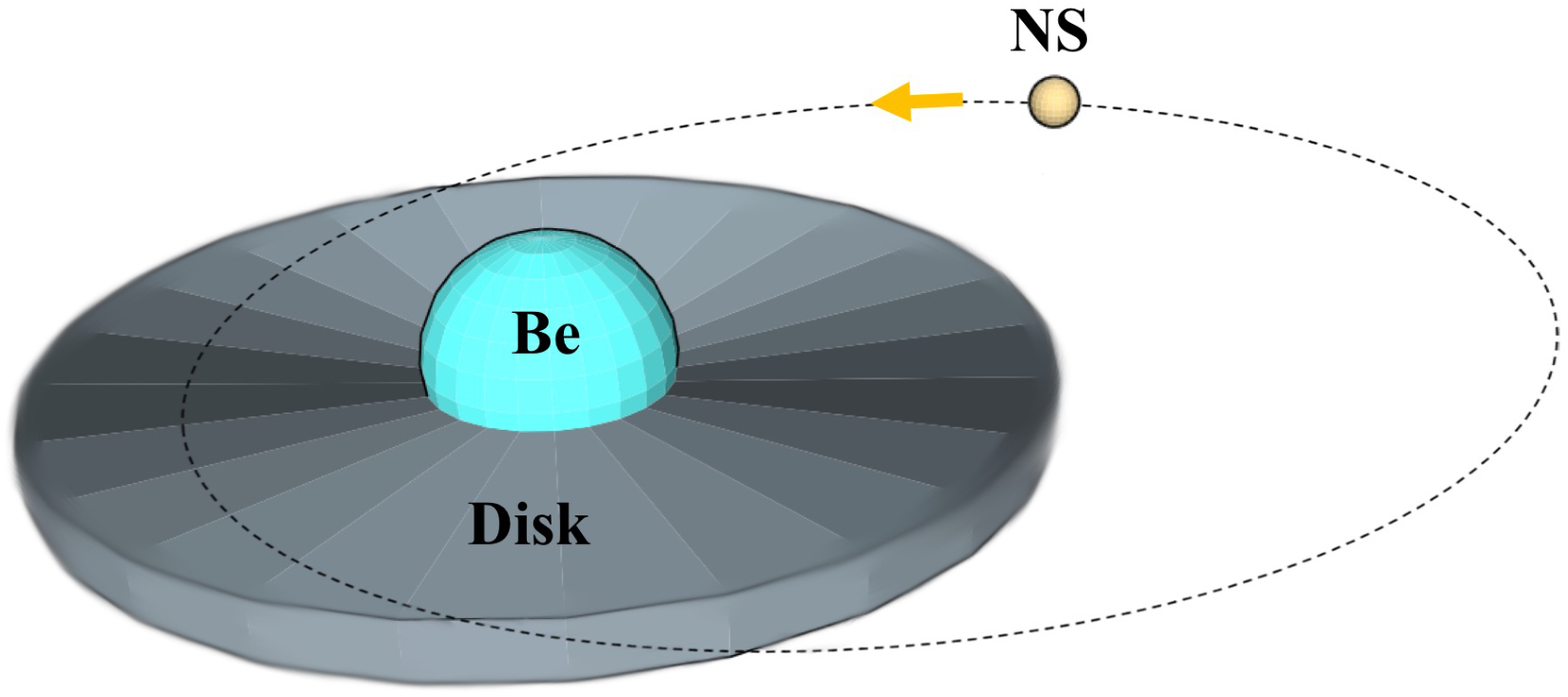}}
\subfigure[]{
\label{Fig.sub.2}
\includegraphics[width=0.56\textwidth]{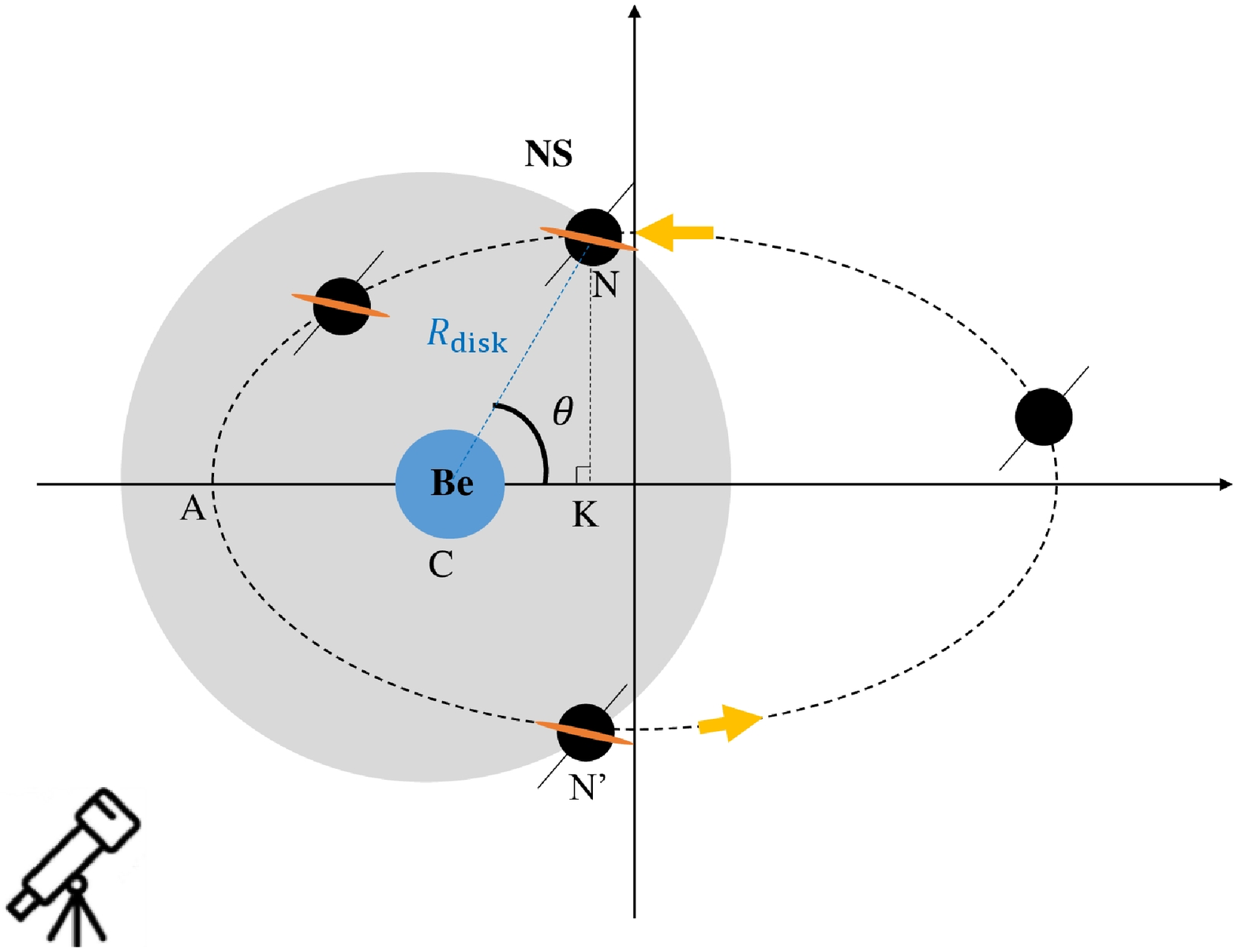}}
\caption{Schematic illustration of the BeXRB model in this work:
(a) the three-dimensional structure of the system with a Keplerian disk of the Be star, and the NS is coplanar with the disk;
(b) the top view of the system. The dotted ellipse represents the orbit of the NS.
The gray area represents the disk of the Be star.
In the disk accretion process, an accretion disk (the orange area) would form around the NS when it is in the circumstellar disk of the Be star and begin to fade during it is out \citep[e.g.,][]{oka01}.
 }
\label{model orbit}
\end{figure}

In a BeXRB system (Figure~\ref{Fig.sub.1}), the NS will accrete material when it is in the disk of the Be star \citep[e.g.,][]{oka01, wil18}.
The interaction between accreted material and NS magnetosphere will change the spin period of the NS.
In this scenario, FRBs might be triggered by starquakes due to the spin evolution of the NS. Different from the model that FRBs are triggered by magnetic stress evolution in magnetar crust \citep{yan21}, for an NS with relatively weaker magnetic field and faster spin, the crust fracturing would be adjusted by the balance between crust stress and centrifugal force.
Because an NS may spin up or spin down during the interaction between its magnetosphere and the accreted material around the NS, the balance of the crust would be destroyed.
The stress of the crust reaching a critical value will cause the crust fracturing.
When the NS moves outside the Be disk, it will spin down by magnetic dipole radiation or gravitational wave radiation.
In the following discussion, we will estimate the rates and energies of the starquakes when the NS is in/out the Be star disk, and study whether accretion-induced starquakes can provide enough energy and explain the high rate of repeating FRBs.

\subsection{Repeating FRBs triggered by NS starquakes in a BeXRB system}\label{sec2.1}

First, we make a discussion about starquakes of the NS induced by the spin evolution.
When the spin period of the NS changes, the oblateness ($\varepsilon$) and the moment of inertia ($I$) will change to readjust the stellar shape. The oblateness of the NS is defined as $\varepsilon \equiv |I - I_0|/I $, where the non-rotating moment of inertia of the NS is taken as $I_0 = 10^{45}\,\unit{g\,cm^{2}}$. The rigidity of the NS resists this change and this corresponding energy is defined as the strain energy
$E_{\text{strain}} = \mathcal{B}\left(\varepsilon-\varepsilon_{0}\right)^{2}$,
where $\mathcal{B}$ is the coefficient and $\varepsilon_{0}$ is a reference oblateness without strain energy.
The mean stress is
\begin{eqnarray}
\sigma=\left|\frac{1}{V} \frac{\partial E_{\text {strain }}}{\partial \varepsilon}\right|=\mu\left|\varepsilon - \varepsilon_{0} \right| ,
\label{sigma}
\end{eqnarray}
where $V$ is the volume of the NS, and $\mu = 2\mathcal{B}/V$ is the mean shear modulus~\footnote{The shear modulus of the crust of the NS can be represented by \citep[e.g.,][]{tom95, dou01, pio05} $ \mu \simeq 6.0 \times 10^{30} \,\unit{ erg\,cm^{-3} }  \rho_{\text{n}}^{4 / 3}(Z/50)^{2}  (615/A)^{4 / 3}[(1-X_{\text{n}})/0.4]^{4 / 3} $, where $\rho_{\text{n}}=2.8\times10^{14}~{\rm g~cm^{-3}}$ is the nuclear density, $Z$ is the number of protons per ion, $A$ is the number of nucleons in a nucleus and $X_{\text{n}}$ denotes the fraction of neutrons outside nuclei.
Because the shear modulus of the NS crust during the accretion process is a little bit softer \citep[e.g.,][]{hae08, zdu08}, here we adopt $\mu \sim 10^{31} \,\unit{ erg\,cm^{-3} }$ in this process.} 
of the NS.
There is a critical stress value $ \sigma_{\text{c}}$, above which, i.e., $\sigma > \sigma_{\text{c}}$, a starquake can take place.

The total energy of the NS can be estimated by \citep{bay71, shap86}
\begin{align}
E &=E_{\text{grav}} + E_{\text{rot}} + E_{\text{strain}}   \notag \\ &=E_{0}+\mathcal{A} \varepsilon^{2}+\mathcal{L}^{2} /(2 I)+\mathcal{B}\left(\varepsilon-\varepsilon_{0}\right)^{2},
\end{align}
where $E_{\text{grav}} = E_{0} + \mathcal{A} \varepsilon^{2}$ is the gravitational energy of the rotating NS, with the energy of the nonrotating star $E_{0} = -3M_{\text{NS}}^2G/(5R_{\text{NS}})$ and the coefficient $ \mathcal{A} = -1/5 E_{0} $ for a self-gravitating incompressible sphere, $E_{\text{rot}} = \mathcal{L}^{2} /(2 I)$ is the rotation energy, with the stellar angular momentum $\mathcal{L} = I \Omega$ and the angular velocity of the NS $ \Omega = 2\pi/P_{\text{NS}}$.
By minimizing the total energy $E$, $\varepsilon$ satisfies
\begin{eqnarray}
\begin{aligned} \varepsilon=&\frac{\Omega^{2}}{4(\mathcal{A}+\mathcal{B})} \frac{\partial I}{\partial \varepsilon}+\frac{\mathcal{B}}{\mathcal{A}+\mathcal{B}} \varepsilon_{0} \\  = & \frac{I_{0} \Omega^{2}}{4(\mathcal{A}+\mathcal{B})}+\frac{\mathcal{B}}{\mathcal{A}+\mathcal{B}} \varepsilon_{0},
\label{epsilon}
\end{aligned}
\end{eqnarray}
where $\partial I(\varepsilon) / \partial \varepsilon=I_{0}$.
By setting $\mathcal{B} = 0$, the reference oblateness is
$\varepsilon_{0}=I_{0} \Omega_{0}^{2} /(4 \mathcal{A})$,
where $\Omega_{0}$ is the angular velocity of the NS with completely unstressed crust.

When the stress $\sigma$ in Eq.(\ref{sigma}) reaches the critical stress $\sigma_{\text{c}}$, a starquake is produced.
The reference oblateness as a reference point shifts $\Delta \varepsilon_{0}$, and thus the actual oblateness of the NS shifts
\begin{eqnarray}
\Delta \varepsilon = \frac{\mathcal{B}}{\mathcal{A}+\mathcal{B}} \Delta \varepsilon_{0}.
\end{eqnarray}
According to the conservation of the angular momentum, $\Delta \varepsilon = |\Delta \Omega/ \Omega|$, where $\Delta \Omega$ is the difference between the angular velocity after and before the starquake.
Based on the observation of pulsar glitches, $\Delta \varepsilon$ can span several orders of magnitude from $\Delta \varepsilon \approx 10^{-12}$ to $\Delta \varepsilon \approx 10^{-5}$ \citep{esp11}.
\cite{lai18} gives an estimation of $\Delta \varepsilon\sim 10^{-10}$ for the Crab pulsar and $\Delta \varepsilon\sim10^{-11}$ for the Vela pulsar from the observations \citep{esp11}.
Then the stress relieved during a starquake can be represented by
\begin{eqnarray}
|\Delta \sigma | = \mu |\Delta \varepsilon_{0} - \Delta \varepsilon| = \mu \frac{\mathcal{A}}{\mathcal{B}} |\Delta \varepsilon |.
\end{eqnarray}
According to Eq.(\ref{sigma}) and Eq.(\ref{epsilon}), the stress in the crust will start to build up again after starquakes at a rate
\begin{eqnarray}
| \dot{\sigma} |=  \mu | \dot{\varepsilon} |= \frac{\mu I_{0}}{2(\mathcal{A}+\mathcal{B})} \Omega |\dot{\Omega}|,
\end{eqnarray}
where $\dot{\Omega}$ is the change rate of angular velocity of the NS.
The next starquake occurs after a time \citep{has15}
\begin{eqnarray}
t_{\text{s}} \approx \frac{|\Delta \sigma |}{|\dot{\sigma}|} = \frac{2\mathcal{A}(\mathcal{A} + \mathcal{B}) |\Delta \varepsilon|}{\mathcal{B} I_0 \Omega |\dot{\Omega}|},
\label{ts}
\end{eqnarray}
i.e., the waiting time of the NS starquakes, where $\dot{\Omega}$ can be derived by the torque $N$ on the NS, i.e., $N = I_0 \dot{\Omega}$. In the following part, we will derive $N$ on the NS in/out of the Be star disk.

Because the accretion process of the NS in a BeXRB system is still unclear, both disk accretion \citep[e.g.,][]{oka01, xuxiao19} and wind accretion \citep[e.g.,][]{ikh07, sha12} are possible scenarios. 
During the NS is in the disk of the Be star, if the interaction radius between the accreted material and the NS magnetosphere $R_{\mathrm{in}}$ and the radius of the light cylinder $R_{\text{lc}}$ satisfies $R_{\mathrm{in}} < R_{\text{lc}}=c/\Omega  \simeq 4.8 \times 10^7 ( P_{\text{NS}}/0.01\,\unit{s})\,\unit{cm}$, there will exist the interaction between the accreted material and the magnetosphere of the NS. In this situation, an accretion torque $N_{\rm acc}$ would change the NS spin evolution, leading to spin up or spin down~\footnote{We define the corotation radius as $R_{\rm co}$  where the local Keplerian angular velocity of the accreted material equals the angular velocity of the NS.  
If $R_{\rm in} < R_{\rm co}$, i.e., accretor state, the material in the inner boundary would be faster than the rotation velocity of the NS, and flows along the magnetic field lines onto the NS \citep{fra02}, which casues that accretion occurs and accelerates the rotation of the NS by the angular momentum transfer. 
If $R_{\rm co} < R_{\rm in} < R_{\rm lc}$, i.e., propeller state, accretion is forbidden,  because the velocity of the accretion disk is less than the rotation velocity of the NS. The material of the accretion disk is thrown out by centrifugal force, which takes away angular momentum and decelerates the NS.}. 
$N_{\rm acc}$ can be represented by \citep[e.g.,][]{gho79b, lv12}
\begin{eqnarray}
N_{\rm acc} \simeq \pm\dot{M}\left(G M_{\text{NS}} R_{\mathrm{in}}\right)^{1 / 2},
\label{Nacc}
\end{eqnarray}
where $\dot{M}$ is the accretion rate of the NS, ``$+$'' corresponds to the spin-up case, and ``$-$'' corresponds to the spin-down case.
The total torque can be represented by $ N_{\text{tot}} = N_{\rm acc} + N_{\mathrm{rad}} $,
with the radiation torque of the NS to be $N_{\mathrm{rad}}$.
Because $|N_{\text{tot}}| \gg |N_{\mathrm{rad}}|$, the starquakes in both the accretor and propeller states would have the same waiting time~\footnote{Here we only consider the direction of the accretion disk of the NS is the same as that of the NS spin.}.

The interaction radius between the accreted material and the NS magnetosphere $R_{\text{in}}$ can be evaluated by $R_{\text{in}}=\xi R_{\text{A}}$. 
$\xi \sim 0.01-1$ is taken for the disk accretion \citep[e.g.,][]{gho79a, fil17} and $\xi \sim 1$ is taken for the wind accretion \citep[e.g.,][]{lv12}, respectively. 
The Alfv\'en radius $R_{\text{A}} = [\mu_{\text{NS}}^4/(2GM_{\text{NS}}\dot{M}^2)]^{1/7}$ is where the magnetic pressure of an NS is equal to the shock pressure of accreted material, with the accretion rate $\dot{M}$. 
$\mu_{\text{NS}} = B_{\text{NS}}R^3_{\text{NS}}/2$ is the magnetic moment of the NS, with the surface magnetic field of the NS to be $B_{\text{NS}}$.
Thus, $R_{\text{in}}$ is represented by
\begin{align}
R_{\mathrm{in}} \simeq & 4.1\times10^7 \,\unit{cm} \left( \frac{\xi}{0.1}\right) \left(\frac{B_{\text{NS}}}{10^{13}\,\unit{G}}\right)^{4/7} \left(\frac{R_{\text{NS}}}{10^6\,\unit{cm}}\right)^{12/7} \notag \\
 &\times \left(\frac{\dot{M}}{\dot{M}_{\text{Edd}}}\right)^{-2/7}\left(\frac{M_{\text{NS}}}{1.4M_{\odot}}\right)^{-1/7}.
\end{align}
The Eddington luminosity is defined as $ L_{\text{Edd}} \equiv 4\pi G  M_{\text{NS}} m_{\text{p}} c/\sigma_{\text{T}}$,
where $m_{\text{p}}$ is the mass of proton and $\sigma_{\text{T}}$ is the Thompson cross section.
The rate of Eddington accretion is
$\dot M_{\text{Edd}} \simeq 10^{18} \left(R_{\text{NS}}/10^6\,\unit{cm}\right) \,\unit{g\,s^{-1}}$.
For BeXRB system, the X-ray luminosity of the type Type \uppercase\expandafter{\romannumeral1} outbursts is about $10^{36}-10^{37}\,\unit{erg\,s^{-1}}$. If we assume that all the kinetic energy of the accretion matter is converted to the luminosity of type Type \uppercase\expandafter{\romannumeral1} outbursts, i.e.,
$L=1/2 \dot{M} v_{\mathrm{ff}}^{2}=G M_{\text{NS}} \dot{M}/R_{\text{NS}}$,
with the free-fall velocity $v_{\text{ff}}=\sqrt{2GM_{\text{NS}}/R_{\text{NS}}}$,
we can derive that the accretion rate of the NS is about $0.1\dot{M}_{\text{Edd}}$.

According to Eq.(\ref{ts}), the waiting time of two starquakes induced by accretion of the NS in the Be star disk can be represented by
\begin{align}
t_{\text{s}} \simeq  & 2.5  \,\unit{day} \left(\frac{\Delta \varepsilon}{10^{-13}}\right) \left( \frac{P_{\text{NS}}}{0.01\,\unit{s}}\right)  \left(  \frac{\xi}{0.1} \right)^{-1/2}\notag \\ &\times \left(\frac{B_\text{NS}}{10^{13}\,\unit{G}}\right)^{-2/7}
 \left(\frac{\dot{M}}{\dot{M}_{\text{Edd}}}\right)^{-6/7},
\end{align}
with $\mu = 10^{31}  \,\unit{erg\,cm^{-3}}$ \citep[e.g.,][]{tom95, dou01, pio05}, $M_{\text{NS}} = 1.4M_{\odot}$, $R_{\text{NS}} = 1.2\times 10^6\,\unit{cm}$, and $|N_{\text{acc}}| \gg |N_{\mathrm{dip}}|$.
Here, we take $P_{\text{NS}}= 0.01\,\unit{s}$ as a reference value, which is consistent with the observation of SAX J0635+0533 with a spin period $0.0338\,\unit{s}$ \citep{kaa99, cus20}.
If $P_{\text{NS}}= 0.02\,\unit{s}$, $\Delta \varepsilon = 10^{-13}$, $\xi=0.1$, $B_\text{NS}=10^{13}\,\unit{G}$,  $M_{\text{NS}} = 1.4M_{\odot}$, $R_{\text{NS}} = 1.2\times 10^6\,\unit{cm}$ and $\dot{M}=\dot{M}_{\text{Edd}}$, we obtain $R_{\mathrm{in}} \simeq 5.3 \times 10^7 \,\unit{cm} < R_{\mathrm{lc}}\simeq 9.5 \times 10^7\,\unit{cm}$ and $t_{\text{s}} \simeq 4.9\,\unit{day}$.
Thus, in an active window of FRB 180916B, there would be several starquakes, leading to several FRBs occurring in the active window.
If $\xi = 0.5$, $P_{\text{NS}}= 0.01\,\unit{s}$, $B_\text{NS}=10^{13}\,\unit{G}$,  $M_{\text{NS}} = 1.4M_{\odot}$, $R_{\text{NS}} = 1.2\times 10^6\,\unit{cm}$ and $\dot{M}=\dot{M}_{\text{Edd}}$, $R_{\mathrm{in}} \simeq 2.6 \times 10^8 \,\unit{cm} > R_{\mathrm{lc}}\simeq 4.8 \times 10^7\,\unit{cm}$, without interaction between the accreted material and the magnetosphere of the NS, the NS will spin down through radiation which will be discussed later.

When the crust of the NS cracks, the magnetic field near the surface would be disturbed, and further propagates as Alfv\'en waves into the magnetosphere \citep[e.g.,][]{tom95,kum20,lu20,yan21}. The released energy can be evaluated by
\begin{align}
E_{\text{s}} \sim & \zeta \frac{B_{\text{NS}}^2}{8\pi} 4\pi R^2_{\text{NS}} \Delta R \simeq  5.0\times10^{42} \,\unit{erg} ~\zeta \left(\frac{B_\text{NS}}{10^{13}\,\unit{G}}\right)^{2} \notag \\
 &\times \left(\frac{R_{\text{NS}}}{10^6\,\unit{cm}}\right)^{2} \left(\frac{\Delta R}{10^5\,\unit{cm}}\right),
\end{align}
where $ \Delta R$ is the thickness of the crust and $\zeta$ is the magnetic energy conversion factor.
Therefore, the energy released by the magnetic energy in the NS crust is enough to produce FRBs.

When the NS is outside the disk of the Be star or $ R_{\rm in} > R_{\rm lc}$, the accretion will be very weak. 
Taking $|\dot{\Omega}|= 2 \pi P_{\text{NS}}^{-2}\dot{P}_{\text{NS}}$ into Eq.(\ref{ts}), the interval between two starquakes due to the radiation of the NS can be estimated by
\begin{align}
t_{\text{s}}' \simeq 2774  \,\unit{day}  \left(\frac{\Delta \varepsilon}{10^{-13}}\right) \left( \frac{P_{\text{NS}}}{0.01\,\unit{s}}\right)^3 \left(\frac{\dot{P}_\text{NS}}{10^{-18}\,\unit{s\,s^{-1}}}\right)^{-1} ,\label{waittime2}
\end{align}
where $\dot{P}_{\text{NS}}$ is the period derivative of the NS, with typical value in range $10^{-20}\,\unit{s\,s^{-1}}$ to $10^{-10}\,\unit{s\,s^{-1}}$ \citep{tay93},
The other parameters are taken as $\mu = 10^{31}  \,\unit{erg\,cm^{-3}}$, $M_{\text{NS}} = 1.4M_{\odot}$, and $R_{\text{NS}} = 1.2\times 10^6\,\unit{cm}$.
This interval is much longer than the orbit period of the BeXRB system, thus it is unlikely to produce radio bursts when the NS moves outside the disk of the Be star.

\begin{figure}[htbp]
\centering
\includegraphics[angle=0,scale=0.55]{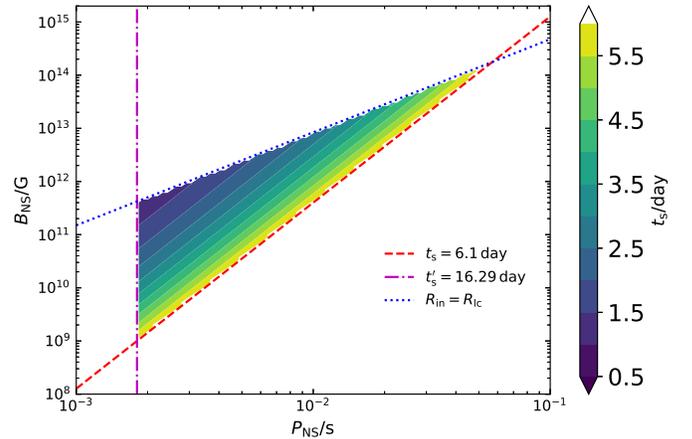}
\caption{The period and magnetic field of the NS in a BeXRB system. The red dashed line, purple dash-dotted line and blue dotted line denote $t_{\text{s}} = 6.1\,\unit{day}$, $t_{\text{s}}' = 16.29\,\unit{day}$ and $R_{\mathrm{in}} = R_{\mathrm{lc}}$, respectively.
Thus, the enclosed area is the range that meets $t_{\text{s}} < 6.1\,\unit{day}$, $t_{\text{s}}' > 16.29\,\unit{day}$ and $R_{\mathrm{in}} < R_{\mathrm{lc}}$.
The darker the color, the shorter the $t_{\text{s}}$.
Model parameters are taken as $\xi = 0.1$, the shear modulus of the crust $\mu = 10^{31}  \,\unit{erg\,cm^{-3}}$, the oblateness of the NS $\Delta \varepsilon = 10^{-13}$,  the accretion rate $\dot{M} = \dot{M}_{\text{Edd}}$, the period derivative of the NS $\dot{P}_\text{NS} = 10^{-18}\,\unit{s\,s^{-1}}$, the mass of the NS $M_{\text{NS}} = 1.4M_{\odot}$, and the radius of the NS $R_{\text{NS}} = 1.2\times 10^6\,\unit{cm}$.}
\label{P-B}
\end{figure}

In Figure~\ref{P-B}, we plot the parameter space of period and magnetic field of the NS. The area enclosed by the red dashed line, purple dash-dotted line and blue dotted line satisfies $t_{\text{s}} < 6.1\,\unit{day}$, $t_{\text{s}}' > 16.29\,\unit{day}$ and $R_{\mathrm{in}} < R_{\mathrm{lc}}$, which is suitable to explain FRB 180916B period activity.
In this area, different colors denote different $t_{\text{s}}$. The parameters are taken as $\mu = 10^{31}  \,\unit{erg\,cm^{-3}}$, $\xi = 0.1$, $\Delta \varepsilon = 10^{-13}$,  $\dot{M} = \dot{M}_{\text{Edd}}$, $\dot{P}_\text{NS} = 10^{-18}\,\unit{s\,s^{-1}}$, $M_{\text{NS}} = 1.4M_{\odot}$, and $R_{\text{NS}} = 1.2\times 10^6\,\unit{cm}$.

When the crust fractures, Alfv\'en waves would be generated at the surface of the NS by the starquake and propagate into the magnetosphere, further produce FRBs at a large radius \citep{kum20, lu20, yan21}. About tens of NS radius, the plasma density in the magnetosphere is not enough to screen the parallel electric field associated with the Alfv\'en wave, leading to a charge starved region.
In this scenario, FRBs could be emitted by bunching coherent curvature radiation due to two-stream instability \citep{kum20, lu20,yan20} or coherent plasma radiation \citep{yan21}.

\subsection{Periodicity and active window of FRB 180916B}\label{sec2.2}

We assume that the disk of the Be star is coplanar with the orbit of the NS for simplify, and FRBs are produced when the NS passes through the disk of the Be star. In Figure \ref{Fig.sub.2}, the Be star is at the left focus of the orbit.
The NS moves counterclockwise.
The point $N$ is where the NS starts to move into the disk of the Be star, and the point $N'$ is where the NS is beginning to move outside of the disk of the Be star.
We define
\begin{eqnarray}
\Delta S_{\text {NS}} (\theta_1, \theta_2) \equiv \frac{1}{2}\int_{\theta_1}^{\theta_2}\left[\frac{a\left(1-e^{2}\right)}{1-e \cos \theta}\right]^{2} d \theta,
\end{eqnarray}
where $\theta$ is the azimuth angle of the NS relative to the Be star, $a$ is the length of semi-major axis of the orbit, and $e$ is the orbital eccentricity.
The area swept out by a line joining the Be star and the NS from the point $N$ to the point $N'$ can be calculated by
$\Delta S_{\text {NS}} (\theta_0, 2\pi-\theta_0)$,
where $\theta_0$ is the initial azimuth angle at the point $N$.
Setting the phase of the orbit periastron as 0.5, the phase at any azimuth angle, $\theta_x$, of the orbit can be represented by
$\phi (\theta_x) = \Delta S_{\text {NS}} (\theta_0, \theta_x)/S_{\text {NS}} - \Delta S_{\text {NS}} (\theta_0, \pi)/S_{\text {NS}} + 0.5$.

According to the Kepler's third law, the length of semi-major axis of the orbit should meet 
$a = \left[ G(M_{\star} + M_{\text{NS}})P_{\text{orb}}^2/4\pi^2 \right]^{1/3} \simeq 4.8\times 10^{12} \,\unit{cm}$,
where the orbit period is $P_{\text{orb}} = 16.29 \,\unit{day}$, the mass of the Be star is taken as $M_{\star} = 15 M_{\odot} $ and the mass of the NS is $M_{\text{NS}} = 1.4 M_{\odot} $.
The total area of the orbit is $S_{\text {NS}}=\pi\left(1-e^{2}\right)^{1 / 2} a^{2}$. 
We define the active fraction of the period as the duty cycle.
According to Kepler's second law, the duty cycle can be represented by
\begin{align}
\mathcal{D} \equiv \frac{\Delta P_{\text {orb}}}{P_{\text {orb}}} = \frac{\Delta S_{\text {NS}}(\theta_0, 2\pi-\theta_0)}{S_{\text {NS}}}=\frac{6.1\,\unit{day}}{16.29\,\unit{day}}\simeq 0.37,
\end{align}
with the active window of FRB 180916B to be $\Delta P_{\text {orb}}$.

\begin{figure}[htbp]
\centering
\includegraphics[angle=0,scale=0.50]{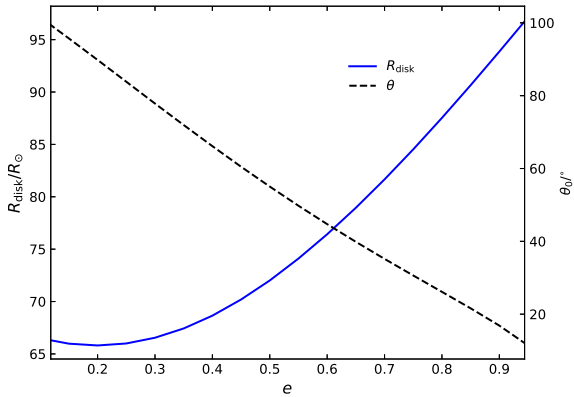}
\caption{The radius of the disk of the Be star $R_{\text{disk}}$ and the initial azimuth angle of the NS just entering the disk $\theta_0$ as functions of the orbital eccentricity of the BeXRB $e$.  
The parameters are taken as $P_{\text{orb}} = 16.29 \,\unit{day}$ and $\Delta P_{\text {orb}} = 6.1 \,\unit{day}. $}
\label{Rdisk}
\end{figure}

The abscissa of the point $N$ satisfies $x_{\text{K}} = (-a^2+aR_{\text{disk}})/c_{\text{e}}$ and $x_{\text{K}}  = R_{\text{disk}} \cos(\theta_0) - c_{\text{e}}$, with $c_{\text{e}}$ the focal length of the orbit and $R_{\text{disk}}$ the radius of the disk. 
For a given orbit period and active window, $R_{\text{disk}}$ and  $\theta_0$ are functions $e$, respectively,
as shown in Figure~\ref{Rdisk}. The larger the eccentricity $e$, the larger the disk radius $R_{\text{disk}}$ and the smaller the initial azimuth angle $\theta_0$ are required.

Next, we establish a three-dimensional model to describe the density distribution of the Be star disk.
The disk of the Be star is usually a Keplerian disk \citep[e.g.,][] {lee91, oka91, qui97, riv13}.
The velocity at the horizontal distance $r$ to the central star is $v = \sqrt{GM_{\star}/r}$, where $G$ is Newton’s gravitational constant. In the $z$ direction vertical to the disk, the gas pressure is balanced with the gravitational force, i.e.,
\begin{eqnarray}
\frac{1}{\rho}\frac{\partial P}{\partial z} = -\frac{GM_{\star}z}{r^3},
\end{eqnarray}
where $\rho$ and $P$ are the density and pressure of the matter, respectively. The height $H(r)$ at radius $r$ is estimated by
\begin{eqnarray}
H(r) \simeq c_{\mathrm{s}}\left(\frac{r}{G M_{\star}}\right)^{1 / 2} r,
\label{Hr}
\end{eqnarray}
with the isothermal sound speed $c_{s}=(k T/\mu_{\text{m}} m_{\mathrm{H}})^{1 / 2}$, where $\mu_{\text{m}} = 0.61$ is the mean molecular weight for a solar composition, $T$ is the isothermal electron temperature and $m_{\mathrm{H}}$  is the mass of hydrogen.
The temperature of the disk is roughly constant, $T \sim 2\times 10^4 \,\unit{K}$, within $100R_{\star}$ in the radial direction \citep[e.g.,][]{mil98, mil99}.
The density structure of the Be star disk can be represented radially as a power law and vertically with an exponential decay, \citep[e.g.,][]{mcg13}, i.e.,
\begin{eqnarray}
\rho(r, z)=\rho_{0}\left(\frac{r}{R_{\star}}\right)^{-\eta} e^{-(z / H(r))^{2}},
\label{diskdensity}
\end{eqnarray}
where the base density $\rho_0$ for most of Be stars ranges from $10^{-13}\,\unit{g\,cm^{-3}}$ to $10^{-10}\,\unit{g\,cm^{-3}}$ \citep[e.g.,][]{car07, jon08, tyc08} and the index $\eta$ ranges from $2$ to $5$ \citep[e.g.,][]{wat87, dou94}

In order to calculate the optical depth of a burst along a certain propagation direction, we first define the direction vector of the line of sight. In the reference frame centered on the Be star, the direction vector of the line of sight can be expressed as $(\cos(\beta_0) \cos(\alpha_0),  \cos(\beta_0) \sin(\alpha_0),  \sin(\beta_0))$, with direction angle $\alpha_0$ and vertical angle $\beta_0$. When the NS moves in the disk of Be star, the optical depth of free-free absorption along the line of sight is calculated by
\begin{eqnarray}
\tau_{\nu} = \int  \kappa_{\nu} d l,\label{optdep}
\end{eqnarray}
with the attenuation coefficient \citep[e.g.,][]{dra11}
\begin{align}
\kappa_{\nu} &\approx 1.772 \times 10^{-26} T_{4}^{-1.5} \nu_9 ^{-2}Z_{i}^2 g_{\text{ff}} n_{e} n_{i} \,\unit{cm}^{-1} \notag\\
&\approx 1.091 \times 10^{-25} Z_{i}^{1.882} T_{4}^{-1.323} \nu_{9}^{-2.118} n_{e} n_{i} \,\unit{cm}^{-1},
\end{align}
where $T=T_410^4\,\unit{K}$ is the temperature, $\nu = \nu_9 10^9 \,\unit{Hz}$ is the frequency, $Z_i \sim 1$ is the charge of ions, $g_{\text{ff}}(\nu, T)$ is the Gaunt factor for the free-free emission, and $n_i \approx n_e$ is assumed with the ion density $n_i$ and electron density $n_e= \rho/\mu_{\rm e} m_{\rm H}$, with the mean molecular weight per electron $\mu_{\rm e}=1.2$ for a solar composition. 
If $\tau_\nu>1$, the observed flux of an FRB would be significantly reduced by free-free absorption.
According to Eq.(\ref{optdep}), in the following part, we will calculate the burst rate at different frequencies from a certain propagation path after the free-free absorption.

When the NS is in the Be star disk, it will accrete the material and trigger FRBs by starquakes as discussed above in Section \ref{sec2.1}.
We assume that the burst rate during the active window satisfying a Gaussian distribution as a function of phase (also see the observation result of Figure~4 in \cite{pas20}). The burst rate reaches the maximum near the periastron, and decreases when the NS is away from the periastron, because the accretion rate depends on the density of the Be star disk.
Such a Gaussian burst rate is also consistent with pulse shapes of Type \uppercase\expandafter{\romannumeral1} outbursts of BeXRBs \citep[e.g.,][]{wil02,bay08, wil08}.
The burst rate at $1\,\unit{GHz}$ as a function of phase can be represented by
\begin{eqnarray}
\mathcal R_{{\rm{1 GHz}}}({\rm{\phi}})=\mathcal R_0\frac{1}{\sigma_{\phi} \sqrt{2 \pi}} e^{-\frac{(\phi -\phi_0)^{2}}{2 \sigma_{\phi}^{2}}},
\end{eqnarray}
where $\mathcal R_0$ is the normalized constant, $\sigma_{\phi}$ is the standard deviation, and $\phi_0 =0.5$. 
According to Extended Figure 1 of \cite{pas20}, for bursts of the same fluence, FRB 180916B is around over an order of 
magnitude more active at $150\,\unit{MHz}$ than at $1.4\,\unit{GHz}$. 
Thus, the ratio of burst rates in $1.4\,\unit{GHz}$ and $150\,\unit{MHz}$ is $\mathcal R_{1.4\,\unit{GHz}}/\mathcal R_{150\,\unit{MHz}} = (1.4\,\unit{GHz}/150\,\unit{MHz})^{-1}$. 
Furthermore, we assume that the burst rate at different frequencies can be represented by
\begin{eqnarray}
\mathcal R_\nu(\phi)=\mathcal R_{{\rm{1 GHz}}}(\phi)\left(\frac{\nu}{\rm{1 GHz}}\right)^{\alpha},
\end{eqnarray}
with $\alpha \approx -1$ as a typical value.

We further generate a sample of FRBs with different frequencies at different phases. The fluences of FRBs, $f_{\nu}(\phi)$, at a phase $\phi$ are assumed to satisfy a normal distribution of $N(\mu_{f_{\nu}}, \sigma_{f_{\nu}}^2)$, where $\mu_{f_{\nu}}$ is the average fluence and $ \sigma_{f_{\nu}}$ is the standard deviation.
According to Extended Figure 1 of \cite{pas20}, the fluence of the low-frequency bursts is higher than that of high-frequency bursts. The average fluence at $1.4\,\unit{GHz}$ of FRB 180916B is a few Jy ms, however at  $150\,\unit{MHz}$ is a few hundred Jy ms.
We assume the average fluence at different frequencies $\nu$ as a power law, i.e.,
\begin{eqnarray}
\mu_{f_{\nu}} = \mu_{f_{1\,\unit{GHz}}}\left(\frac{\nu}{\rm{1 GHz}}\right)^{\gamma},
\end{eqnarray}
where $\mu_{f_{1\,\unit{GHz}}}$ is the average fluence at $1\,\unit{GHz}$, and $\gamma$ is the power law index with $\gamma \approx -2$ as a typical value.

Involving the absorption of the disk of the Be star, the observed fluence of each burst at phase $\phi$ can be estimated by
\begin{eqnarray}
f_{\text{$\nu$,obs}}(\phi) = f_{\nu}(\phi) e^{- \tau_{\nu}(\phi)}.
\end{eqnarray}
The observed burst rate $\mathcal R_{\nu,\text{obs}}$ is determined by the fluence threshold of a specific telescope. When $f_{\text{$\nu$,obs}}(\phi)$ of a burst is greater than the threshold, it is regarded as observable.
The thresholds of Apertif, CHIME and LOFAR are about $1 \,\unit{Jy\,ms}$ \citep{pas20}, $5.3 \,\unit{Jy\,ms}$ \citep{Chime/Frb2020}, $50 \,\unit{Jy\,ms}$ \citep{pas20}, respectively.
Next, in order to explain the activity window of FRB 180916B, we will simulate the observable active window of FRBs affected by the free-free absorption of the disk of the Be star.

\begin{figure}[htbp]
\centering
\includegraphics[angle=0,scale=0.55]{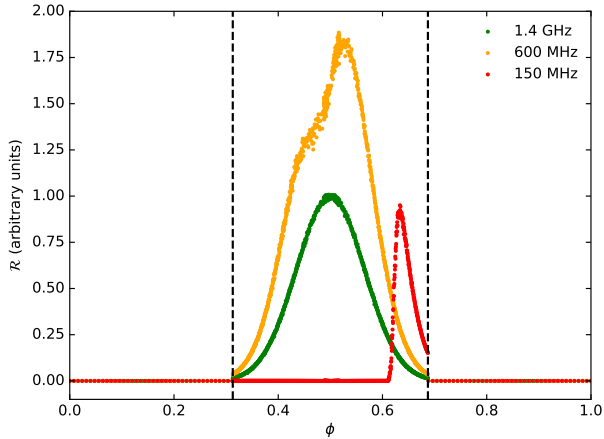}
\caption{The simulation result of the burst rate with phase due to free-free absorption of the Be star disk.
The burst rate is plotted in an arbitrary units.
The dotted green, orange, and red lines denote bursts rate at $1.4\,\unit{GHz}$, $600\,\unit{MHz}$, and $150\,\unit{MHz}$, respectively.
The range between the two dashed black lines indicates that the NS is in the Be star disk and it is the active window of bursts in the BeXRB system.
The parameters are taken as $R_{\star} = 5 R_{\odot}$,  $M_{\star} = 15 M_{\odot}$, $T = 2\times10^4\,\unit{K}$, $\eta = 4.5$, $\rho_0 = 1.6\times10^{-13}\,\unit{g\,cm^{-3}}$, $e=0.35$, $\alpha = -1.0$, $\sigma_{\phi} = 0.065$, $\mathcal{R}_0 = 100\,\unit{h^{-1}}$, $\mu_{f_{1\,\unit{GHz}}} = 5 \,\unit{Jy\,ms}$, $ \sigma_{f_{\nu}} = 0.3 \mu_{f_{\nu}}$, $\gamma = -2.0$, $\alpha_0 = -0.7\pi$ and $\beta_0 = -0.05\pi$.}
\label{ER}
\end{figure}

In general, the radius of Be stars, $R_{\star}$, ranges from $\sim 3R_{\odot}$ to $\sim13R_{\odot}$ and the radius of the Be star disk, $R_{\text{disk}}$, ranges from about $1.5R_{\star}$ to $17R_{\star}$ \citep{riv13}.
The mass of the Be star is $>8M_{\odot}$ \citep{rei11}.
The eccentricity of binary orbits can be 0.1 to 0.9 \citep{zio02}.
If the central star is a B0 main-sequence star, the typical parameters are $R_{\star} = 7.41R_{\odot}$, $M_{\star} = 17.8 M_{\odot}$ \citep{all73}.
In order to explain the observation properties of the frequency-dependent active window of FRB 180916B, we find that the following parameters are appropriate: the radius of Be star is $R_{\star} = 5 R_{\odot}$, the mass of the Be star is $M_{\star} = 15 M_{\odot}$, the temperature of the disk is $T = 2\times10^4\,\unit{K}$, the index of the density structure of the disk is $\eta = 4.5$, the base density of the density structure is $\rho_0 = 1.6\times 10^{-13}\,\unit{g\,cm^{-3}}$, the orbital eccentricity is $e=0.35$, the standard deviation of the burst rate is $\sigma_{\phi} = 0.065$,  the reference burst rate is $\mathcal{R}_0 = 100\,\unit{h^{-1}}$,
the index of the burst rate with frequency is $\alpha = -1.0$,
the average fluence of bursts at $1\,\unit{GHz}$ is $\mu_{f_{1\,\unit{GHz}}} = 5 \,\unit{Jy\,ms}$,
the standard deviation of fluence of bursts is $ \sigma_{f_{\nu}} = 0.3 \mu_{f_{\nu}}$,
the power law index of the average fluence with frequency is $\gamma = -2.0$,
and the direction of the burst is $\alpha_0 = -0.7\pi$ and $\beta_0 = -0.05\pi$. Then the disk of the Be star is about $67.4R_{\odot}$. The frequency-dependent burst rate with phase is shown in Figure~\ref{ER}.
We can see that the peaks of active windows of high-frequency and low-frequency bursts appear at different phases. Low-frequency bursts are observed later than the high-frequency in an orbital cycle.
Bursts with frequency of $600\,\unit{MHz}$ have wider periodic activity window.
The dotted red line denotes the burst rate at $150\,\unit{MHz}$. Because of the absorption, it has the greatest influence on the $150\,\unit{MHz}$ bursts and the burst rate at $600\,\unit{MHz}$ does not follow the shape of Gaussian model. The above results are consistent with the observation properties of FRB 180916B.

\subsection{The dispersion measure contribution from the Be star disk}
\label{sec2.3}
When the NS is in the Be star disk, the observed DM will also change with orbit phase.
In general, the total observed DM can be divided into \citep{deng14, zhang18}
${\rm DM_{tot}} = {\rm DM_{MW}} + {\rm DM_{IGM}} +  ({\rm DM_{host}} + {\rm DM_{src}})/(1+z)$,
where $\rm{DM_{MW}}$ and $ \rm{DM_{IGM}}$ are the contributions from the Milky Way and the intergalactic medium, respectively, $\rm{DM_{host}}$ and $\rm{DM_{src}}$ are the contributions from the host galaxy and source local environment in the rest frame of the FRB, respectively, and $z$ is the redshift of the FRB. 
\cite{yang17} investigated the various possibilities that may cause DM variations of an FRB and finds that only the near-source plasma can cause observable DM variations.
In a BeXRB system, we mainly consider the DM variations contributed from $\rm{DM_{src}}$, i.e., $\rm{DM_{disk}}$, the dispersion measure contribution from the Be star disk.

\begin{figure}[htbp]
\centering
\includegraphics[angle=0,scale=0.5]{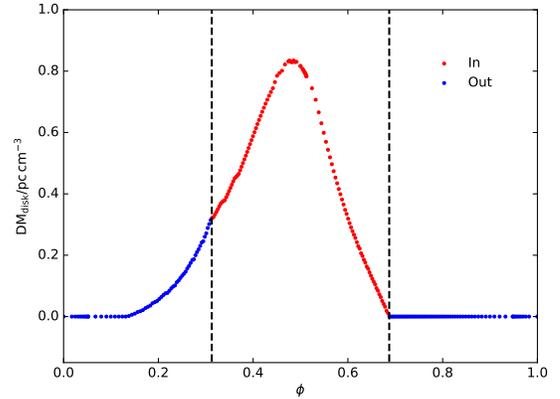}
\caption{The dispersion measure contribution of the Be star disk at different phases. 
The range between the two dashed black lines denotes that the NS is in the Be star disk.
The dotted red and blue lines denote $\rm{DM_{disk}}$ when the NS is in/outside the Be star disk, respectively.
The taken parameters are the same with Figure~\ref{ER}, i.e., $R_{\star} = 5 R_{\odot}$,  $M_{\star} = 15 M_{\odot}$, $T = 2\times10^4\,\unit{K}$, $\eta = 4.5$, $\rho_0 = 1.6\times10^{-13}\,\unit{g\,cm^{-3}}$, $e=0.35$, $\alpha_0 = -0.7\pi$ and $\beta_0 = -0.05\pi$.}
\label{DM}
\end{figure} 

Different from an observable increasing DM of FRB 121102 \citep[e.g.,][]{oos20, zhao21}, there is no obvious DM variation with phase for FRB 180916B, and the largest DM variation is about $ 4 \,\unit{pc\,cm^{-3}}$ \citep[see top panel of Extended Figure 7 in][]{pas20}.  
Based on the density profile of the Be star disk, Eq.(\ref{diskdensity}), we obtain $\rm{DM_{disk}}$ at different phases, as is shown in Figure~\ref{DM}. 
The parameters are taken the same with Figure~\ref{ER}.
The red part corresponds to the DM in active window, and the blue part corresponds to the DM in the non-active window. In non-active window, since the NS moves out of the Be star disk, the burst rate is much lower according to Eq.(\ref{waittime2}.), although it also appears DM variation at some certain line-of-sight directions.
For the given parameters satisfying the properties of the observed frequency-dependent active window of FRB 180916B,
we find that the dispersion measure contribution from the Be star disk is  $< 1 \,\unit{pc\,cm^{-3}}$. 

\section{Discussion and Conclusions}\label{sec3}

Recently, FRB 180916B was found to appear frequency-dependent periodic activities with a period of 16.35 day \citep{Chime/Frb2020}. Such a period might correspond to the orbit period of a binary system \citep{dai16, sma19, iok20, lyu20, dai20b, gu20, mot20, dec21, kue21, den21, wad21}.
FRB 180916B is $\sim250~{\rm pc}$ offset from the brightest region of the nearest young stellar clump in its host galaxy. The large offset is in tension with scenarios that invoke young magnetars born from core collapse supernovae, but is well consistent with the case of the HMXBs.

In this work, we propose that a BeXRB system might be the source of a repeating FRB with periodic activities.
When the NS accretes the material of the Be star disk, the interaction between the NS magnetosphere and the accreted material results in the spin evolution of the NS, leading to the stress in the NS crust changed.
When the stress of the crust reaches the critical value, a starquake occurs, and further produces FRBs.
The Alfv\'en waves are launched by starquakes, and convert to coherent radio emission at a distance of a few tens of NS radius \citep{kum20,lu20,yan21}.
The interval between starquakes is estimated to be a few days.
When the NS is not in the disk of the Be star, due to the long waiting time, the radio bursts occurring at the non-active phase would be too rare to be observed.
With the observed burst rate of FRB 180916B, we can constrain the period and magnetic field of the NS to limited parameter space.

On the other hand, we study the frequency-dependent active window of FRB 180916B in the above scenario. When the NS is in the disk of the Be star, the optical depth of free-free absorption along the line of sight would evolve with orbit phase. If the optical depth is larger than unity, the observed flux of an FRB would be significantly reduced.
FRBs with fluences lower than the telescope threshold would not be observable. Since the free-free absorption is frequency-dependent, the observed active window of FRBs would be frequency-dependent naturally.
Based on the simulation of our model, we find that the active windows of high-frequency and low-frequency bursts peak at different phases, as shown in Figure \ref{ER}. Low-frequency bursts are observed later than the high-frequency in an orbital cycle, which is consistent with the observation properties of FRB 180916B.
We also evaluate the DM contribution from the Be star disk.
It is lower than $ 1 \,\unit{pc\,cm^{-3}}$ when the NS is in the Be star disk, which is also consistent with the observation of FRB 180916B.

It is noteworthy that several more parameters could
affect the features of FRBs with periodic activities.
For instance, the waiting time between bursts is inversely proportional to the strength of the magnetic field, the accretion rate and the spin angular
velocity of the NS. Meanwhile, the active window would be 
shorter if the orbital eccentricity is larger for a given semi-major axis. The profile of the burst rate in the active window would also depend on the length of the semi-major axis of the orbit.

In this model involving the BeXRB, repeating FRBs with periodic activities might be produced by the NS crust fracturing adjusted by the balance between crust stress and NS spin.
To satisfy the observed high burst rate of FRB 180916B, i.e., the waiting time of starquakes induced by accretion $ < 6.1\,\unit{day}$ and radiation $ > 16.29\,\unit{day}$, the NS should have a fast spin period of $P_{\rm NS}\sim0.02~{\rm s}$, a relative large surface magnetic field of $B_{\rm NS}\sim 10^{13}~{\rm G}$, and a small oblateness variation of $\Delta \epsilon\sim10^{-13}$. Such an NS in the BeXRB system is allowable but relatively rare, which might explain why most Galactic BeXRBs do not show FRBs.
On the other hand, in the BeXRB scenario, the multiwavelength counterparts of FRBs could be Type \uppercase\expandafter{\romannumeral1} outbursts. However, since most FRBs occurs at cosmological distance, the FRB-associated Type \uppercase\expandafter{\romannumeral1} outbursts would be hard to be detected. For the FRB-associated Type \uppercase\expandafter{\romannumeral1} outbursts with typical luminosity of $10^{36}-10^{37}~{\rm erg~s^{-1}}$, they could be detected within distance $\lesssim ~{30 \,\rm kpc}$ for current main X-ray detectors, such as Insight-HXMT \citep{hxmt}.

\acknowledgements
We thank the anonymous referee for providing helpful comments and suggestions. 
We also thank Qian-Cheng Liu, Bin Hong and Qin Han for helpful discussions. This work was supported by the National Key Research and Development Program of China (grant No. 2017YFA0402600), the National SKA Program of China (grant No. 2020SKA0120300), and the National Natural Science Foundation of China (grants 11833003, U1831207 and 11988101). Y.P.Y is supported by National Natural Science Foundation of China grant No. 12003028 and Yunnan University grant No.C176220100087.

\end{document}